\documentclass[12pt]{article}

\usepackage{amsmath,amssymb,amsfonts}
\usepackage[paper=letterpaper,margin=1.0in]{geometry}
\usepackage{graphicx}

\begin{document} 

\newcommand{\be}{\begin{equation}}
\newcommand{\ee}{\end{equation}}
\newcommand{\bea}{\begin{eqnarray}}
\newcommand{\eea}{\end{eqnarray}}
\newcommand{\ba}{\begin{array}}
\newcommand{\ea}{\end{array}}
\newcommand{\ben}{\begin{enumerate}}
\newcommand{\een}{\end{enumerate}}
\newcommand{\bi}{\begin{itemize}}
\newcommand{\ei}{\end{itemize}}
\newcommand{\bc}{\begin{center}}
\newcommand{\ec}{\end{center}}
\newcommand{\bfig}{\begin{figure}}
\newcommand{\efig}{\end{figure}}
\newcommand{\bq}{\begin{quotation}}
\newcommand{\eq}{\end{quotation}}
\newcommand{\bt}{\begin{table}}
\newcommand{\et}{\end{table}}
\newcommand{\btab}{\begin{tabular}}
\newcommand{\etab}{\end{tabular}}
\newcommand{\bs}{\begin{slide}}
\newcommand{\es}{\end{slide}}
\newcommand{\nn}{\nonumber}
\newcommand{\eref}[1]{(\ref{#1})}

\newcommand{\AdS}[1]{{\rm AdS}_{#1}}
\newcommand{\pa}{\partial}



\begin{center}
{\Large \bf Entropy and Gravitation: From Black Hole \\ 
Computers to Dark Energy and Dark Matter}
\end{center}

\vskip 1.0cm

\centerline{{\bf
Y.\ Jack Ng
\footnote{\tt yjng@physics.unc.edu}
}}

\vskip 0.5cm

\begin{center}
{\it
Institute of Field Physics, Department of Physics and
Astronomy,\\
University of North Carolina, Chapel Hill, NC 27599, U.S.A.
}
\end{center}

\vspace{1cm}

\begin{abstract}

We show that the concept of entropy and the dynamics of gravitation provide the
linchpin in a unified scheme to understand
the physics of black hole computers, spacetime foam, dark energy, dark matter
and the phenomenon of turbulence. 
We use three different methods to estimate the foaminess of spacetime, which, in
turn, provides a back-door way to derive the
Bekenstein-Hawking formula for black hole entropy and the holographic
principle.  Generalizing the discussion
for a static spacetime region to the cosmos,
we find a component of dark energy (resembling an effective 
positive cosmological constant of the correct magnitude)
in the current epoch of the universe.  The conjunction of entropy and 
gravitation is shown to
give rise to a phenomenological
model of dark matter, revealing the natural emergence, in galactic and cluster
dynamics, of a critical acceleration parameter
related to the cosmological constant; the resulting mass profiles are consistent
with observations.  Unlike ordinary matter,
the quanta of the dark sector are shown to obey infinite statistics.  This
property of dark matter may lead to some non-particle
phenomenology, and may explain why dark matter particles have not been detected
in dark matter search experiments.
We also show that there are deep similarities between the problem of 
``quantum
gravity" (more specifically, the holographic spacetime foam) and turbulence.

\end{abstract}

\vspace{1cm}

Keywords: entropy, gravitation, spacetime foam, quantum foam, holography, dark
energy, dark matter, infinite statistics,  turbulence

\newpage

\section{Introduction}

What is the difference between a computer and a black hole? This question is not
a joke,
but is an intriguing problem in modern 
physics.\cite{llo04}  The
reason
can be traced to the fact that all physical systems are computers.
Every elementary particle stores bits of data, and every time two such particles
interact,
those bits are transformed.
Black holes are merely the most exotic example of the general principle that the
universe
registers and processes information.\\

The principle is not new. In the 19th century, the founders of statistical
mechanics developed
what would later be called information theory to explain the laws of
thermodynamics. 
The key player in information theory is entropy $S$, the macroscopic
thermodynamic quantity characterizing
disorder.  The second law of thermodynamics stipulates that disorder as embodied
by entropy always increases. 
Entropy $S$ can be
written in terms of a microscopic probabilistic quantity $W$ as 
$ S = k {\rm log} W$,
where $k$ is the Boltzmann
constant.  Deeply ingrained in probabilities, $S$ finds its true home in quantum
mechanics.
And the confluence of physics and information theory flows from the central
maxim of quantum mechanics: at bottom, nature is discrete.  It is the
quantum-mechanical nature of
information that is responsible for the computational ability of black holes;
without quantum effects,
a black hole would destroy, rather than process, information.\\

Black holes, though exotic, are, in a way, the simplest gravitational systems. 
To examine the properties of black hole
computers, we can start with a more general discussion of aspects of quantum
gravity, the synthesis of quantum
mechanics and general relativity.  If space-time, like every thing else,
undergoes quantum mechanical fluctuations,
then space is composed of an ever-changing arrangement of bubbles which John
Wheeler called spacetime foam,
also known as quantum foam. \cite{Wheeler}  As we will show, quantum
fluctuations of
spacetime determine the precision with which the geometry of spacetime can
be measured, and they limit the
power of computers in general, black hole computers in particular.   Applied to
cosmology, spacetime foam physics
leads to the prediction of a dark energy component in the current Universe (of
the correct magnitude).   Combined with ideas
from gravitational thermodynamics and entropic gravity, we
are led to a phenomenological model of dark matter
in which a critical acceleration parameter, related to the (effective) 
cosmological constant, emerges. \\

This review article on entropy and gravitation is organized as follows:
In section 2.1, we discuss a gedanken experiment to measure the foaminess of 
spacetime, 
more specifically the induced uncertainties in distance (and time) measurements.
In section 2.2 we rederive these results by the method of mapping the geometry 
of spacetime, 
which also provides a way to derive the holographic
principle.  The results for spacetime fluctuations are then applied to the 
discussion of 
black holes in section 3 to deduce black-hole entropy, lifetime, and power as a 
computer.  
The discussion for a static spacetime region with low
spatial curvature in section 2 is generalized, in section 4, to the case of an 
expanding 
universe, uncovering the constituents of dark energy (of the correct magnitude 
in the present 
era of the universe) in the form of (extremely) long-wavelength quanta
which, thus, act like a positive cosmological constant.  In this section we 
also argue how the results found for spacetime 
fluctuations indicate why the universe necessarily contains 
more than ordinary matter. (One may even suggest that quantum gravity, in 
combination with thermodynamics, naturally demands the existence of a dark 
sector.)  
Section 4.2 is used to show that the quanta of dark energy, unlike 
ordinary matter,  
obey an exotic statistics known as infinite statistics (also known as quantum 
Boltzmann 
statistics).  Another method to infer (and to check the consistency of the 
results for) 
spacetime fluctuations and the magnitude of dark energy is given in section 5 
by applying causal set theory and unimodular gravity.
Section 6 is devoted to the construction of a phenomenological dark
matter model (called Modified Dark Matter (MDM))
by generalizing gravitational thermodynamics and entropic gravity 
arguments to 
a spacetime with positive cosmological constant (like ours).  Then we show that 
dark matter quanta (like dark energy) obey infinite statistics, and briefly 
enumerate some of MDM's
quantitative and qualitative successes (so far).  In section 7, we show some 
deep similarities 
between the physics of spacetime foam and turbulence. We give a short conclusion
 in section 8. 
There are two appendices: In Appendix A, we discuss 
energy-momentum 
fluctuations and some possible tests of spacetime foam.  For completeness
we give a short introduction to the subject of infinite statistics in 
Appendix B.\\

On notations, the subscript ``P" denotes Planck units; thus $l_P \equiv (\hbar G
 /c^3)^{1/2} \sim 10^{-33}$ 
cm is the Planck length etc. On units, $k_B$ (the Boltzmann constant) and $\hbar
$ and c are often put equal 
to 1 for simplicity.\\

\section{Quantum Fluctuations of Spacetime}

At small scales, spacetime is fuzzy and foamy due to quantum fluctuations.
One manifestation of the fluctuations is in the induced uncertainties in
any distance measurement.  We will derive the uncertainties or fluctuations
by two independent methods \cite{wigner,llo04} in the following two 
subsections.
\\

\subsection{Gedanken Experiment}
 
Consider the following experiment to measure the distance $l$ between two 
points.
Following Wigner\cite{SW}, we put
a clock at one of the points and a mirror at the other.  By sending a
light signal from the clock to the mirror in a timing experiment, we can
determine the distance.
However, the quantum uncertainty in the positions of
the clock and the mirror introduces an inaccuracy $\delta l$ in the
distance measurement.  Let us concentrate on the clock (of mass
$m$).  If it has a linear
spread $\delta l$ when the light signal leaves the clock, then its position
spread grows to $\delta l + \hbar l (mc \delta l)^{-1}$
when the light signal returns to the clock, with the minimum at
$\delta l = (\hbar l/mc)^{1/2}$.  Hence one concludes that
\begin{equation}
\delta l^2 \gtrsim \frac{\hbar l}{mc}.
\label{sw}
\end{equation}
One can supplement this requirement with a limit from
general relativity\cite{wigner}.  To wit, let the clock be a
light-clock consisting of two mirrors (each of mass $m/2$), a
distance $d$ apart, between which bounces a beam of light.  For
the uncertainty in distance measurement not to be greater than $\delta l$,
the clock must tick off time fast enough so
that $d/c \lesssim \delta l /c$.  But $d$, the
size of the clock, must be larger than the Schwarzschild radius $Gm/c^2$ of
the clock, for otherwise one cannot read the registered time.
From these two conditions, it follows that
\begin{equation}
\delta l \gtrsim \frac{Gm}{c^2},
\label{ngvan}
\end{equation}
the product
of which with Eq.~(\ref{sw}) yields \cite{wigner,Karol}
\begin{equation}
\delta l \gtrsim (l l_P^2)^{1/3} = l_P \left(\frac{l}{l_P}\right)^{1/3}.
\label{nvd1}
\end{equation}
A gedanken experiment to measure a time interval
$T$ gives an analogous expression:
$ \delta T \gtrsim (T t_P^2)^{1/3}.$ \\

\subsection{Mapping the Geometry of Spacetime}

Since quantum fluctuations of spacetime manifest themselves in the form of
uncertainties in the
geometry of spacetime, the structure of spacetime foam can be inferred from the
accuracy with
which we can measure that geometry.  \cite{llo04}  Let us consider a spherical
volume of
radius $l$ over the amount of time $T = 2l/c$ it takes light to cross the
volume.  One way to map out
the geometry of this spacetime region is to fill the space with clocks,
exchanging signals with
other clocks and measuring the signals' times of arrival. This process of
mapping the geometry is a
sort of computation, in which distances are gauged by transmitting and
processing information; hence
the total number of operations
is bounded by the Margolus-Levitin theorem\cite{Lloyd} in quantum computation,
which stipulates
that the rate of operations for any computer cannot exceed the amount of energy
$E$ that is available
for computation divided by $\pi \hbar/2$.   A total mass $M$ of clocks then
yields, via the Margolus-Levitin
theorem, the bound on the total number of operations given by $(2 M c^2 / \pi
\hbar) \times 2 l/c$.  But to
prevent black hole formation, $M$ must be less than $l c^2 /2 G$.  Together,
these two limits imply that the
total number of operations that can occur in a spatial
volume of radius $l$ for a time period $2 l/c$ is no greater than $\sim
(l/l_P)^2 $.  
(Here and henceforth we neglect multiplicative constants of order unity, and set
$c=1=\hbar$.)
To maximize spatial resolution, each clock must tick only once during the entire
time period. 
And if we regard the operations partitioning the spacetime volume into ``cells",
then
on the average each cell occupies a spatial volume no less than $ \sim l^3 / (
l^2 / l_P^2) = l l_P^2 $, yielding an average
separation between neighboring cells no less than $l^{1/3} l_P^{2/3}$.  This
spatial separation is interpreted as the
average minimum uncertainty in the measurement of a distance $l$, that is,
$\delta l \gtrsim l^{1/3} l_P^{2/3}$,
in agreement with the result obtained in the previous subsection.  (We will use
yet another argument to check
this result in section 5.) \\

As an application, we can now heuristically derive the holographic principle.
Since, on the average, each cell occupies
a spatial volume of $l l_P^2$, a spatial region of size $l$ can contain no more
than $l^3/(l l_P^2) = (l/l_P)^2$ cells.  Thus this
spacetime foam model corresponds to the case of maximum number of bits of
information $l^2 /l_P^2$ in a spatial region
of size $l$, that is allowed by the holographic principle 
\cite{wbhts,GPH},
according to which, the maximum amount of
information stored in a region of space scales as the area of its
two-dimensional surface, like a hologram. Accordingly, we will refer to this
spacetime foam model (corresponding to $\delta l \gtrsim l^{1/3} l_P^{2/3}$)
as the holographic spacetime foam model.
\\

\section{Clocks, Computers, and Black Holes}

In this section we will show that the properties of black holes are inextricably
intertwined with those of spacetime.
For example, the strange scaling of space fluctuations with the cube root of 
distances provide a back-door way to derive the Bekenstein-Hawking formula
for black hole memory. \cite{BeHa1,BeHa2}\\

 But let us first consider a clock (technically, a simple and ``elementary''
clock, not composed of smaller
clocks that can be used to read time separately or sequentially, with  
a black hole clock being the limiting example), capable
of resolving time to an accuracy of $t$, for a period of
$T$ (the running time or lifetime of the clock).
Then bounds on the resolution time and the lifetime of the clock can be
derived by following an argument very similar
to that used above in the analysis of the gedanken experiment to measure
distances.
Actually, the two arguments are so similar that one can identify the
corresponding quantities:
\begin{equation}
\delta l /c \leftrightarrow t;  \hspace{.3in}
l/c \leftrightarrow T.
\label{ltoclock}
\end{equation}

It follows that the following limits \cite{SW,PRL} hold:
\begin{equation}
t^2 \gtrsim \frac{\hbar T}{mc^2}, \hspace{.3in}
t \gtrsim \frac{Gm}{c^3}, \hspace{.3in}
T/ t^3 \lesssim t_P^{-2} = \frac{c^5}{\hbar G},
\label{clock}
\end{equation}
which are, respectively, the analogues of Eqs.~(\ref{sw}), 
(\ref{ngvan}) and (\ref{nvd1}). \\

One can easily translate
the relations for clocks given above
into useful relations for a simple computer
(technically, it refers to a computer designed to perform highly serial
computations, i.e., one that is not divided into subsystems computing in
parallel -- like a black hole computer which acts as a single unit).
Let $\nu$ denote
the clock rate of the computer, i.e., the number of operations
per bit per unit time, and $I$ the number
of bits of information in the memory space of a simple computer. Then one can 
identify the corresponding quantities for simple clocks and simple computers as
\begin{equation}
\frac{1}{t} \leftrightarrow \nu;  \hspace{.3in}
\frac{T}{t} \leftrightarrow I.
\label{cltocomp}
\end{equation}

Now we can apply what we have learned about clocks and computers
to black holes.\cite{PRL}
Let us consider using a black hole to measure time.
It is reasonable to use
the light travel time around the black hole's horizon as the resolution
time of the clock,
i.e., $t \sim \frac{Gm}{c^3} \equiv t_{BH}$, then
from the last of Eq.~(\ref{clock}), one immediately finds that
$T \sim \frac{G^2 m^3}{\hbar c^4} \equiv T_{BH}$,
recovering Hawking's result for black hole lifetime!
(Note that the lifetime bound is saturated for black holes.)\\

Applying the results for $T_{BH}$ and $t_{BH}$, we readily find
the number of bits in the memory space of a black hole computer as
$I = \frac{T_{BH}}{t_{BH}} \sim \frac{m^2}{m_P^2} \sim \frac{r_S^2}{l_P^2}$,
where $m_P = \hbar/(t_P c^2)$ is the Planck mass, $m$ and $r_S^2$ denote the
mass and event
horizon area of the black hole respectively.
This gives the number of bits $I$ as the event horizon area in Planck units,
in agreement with
the identification of black hole entropy \cite{BeHa1,BeHa2}.  
(Recall that entropy $S$ and the 
number of bits $I$ are related by $S = k_B I {\rm ln} 2$.) \\

All these results reinforce the conceptual interconnections of the
physics underlying spacetime foam, black holes, and
computation.  It is interesting that all black hole computers obey
the universal relation (obtained by using the computer analogue of the 
last equation in 
Eq.~(\ref{clock})): 
$I \nu^2 \sim  c^5/\hbar G$, which 
mathematically demonstrates the linkage between information and
the theories of special relativity (where the defining parameter is 
c), general relativity (G) and quantum mechanics ($\hbar$).\\

\section{Dark Energy}

We can now apply the insights we have learned from the fine-scale structure of
spacetime to cosmology and
fundamental physics to learn the behavior of cosmic dark energy.  

\subsection{Spacetime Foam and Dark Energy}

As shown in section 2.2 on mapping the geometry of space-time,
maximum spatial resolution (which leads to the 
holographic bound) requires maximum energy
density (that is allowed to avoid the collapse into a black hole) given by 
\begin{equation}
\rho \sim \frac{l/G}{l^3} = (l l_P)^{-2}. 
\label{eq:rho}
\end{equation}
Let us now generalize this discussion for a static spacetime region 
with low spatial curvature to the case of an expanding universe by 
substituting $l$ by $1/H$, where $H$ is the Hubble parameter.
\cite{Arzano,plb}  Eq. (\ref{eq:rho})
yields the cosmic energy density $\rho \sim
 \left(\frac{H}{l_P}\right)^2 \sim (R_H l_P)^{-2}$.
Next, recall that we have also shown that the Universe contains $I \sim (R_H/l_P
)^2$ bits
of information ($\sim 10^{122}$ for the current epoch).\cite{Arzano}  Hence the
average energy carried by each of these bits or quanta is $\rho R_H^3/I \sim 
R_H^{-1}$. These long-wavelength bits or ``particles'' 
carry negligible kinetic energy. (Note the quotations around the word 
``particles". 
Such long-wavelength quanta can hardly be called particles. We will simply call 
them ``particles".)
Since pressure (energy density) is given by kinetic energy minus (plus)
potential energy, a negligible kinetic energy means that
the pressure of the unconventional energy is roughly equal to minus its
energy density, leading to accelerating cosmic expansion,
in agreement with observation \cite{SNa}.
This scenario is very similar to that of quintessence,
but it has its origin in
local small scale physics -- specifically, the holographic spacetime
foam. \cite{holocos,yjng05} 
Thus intriguingly, the large-scale ($\sim R_H$) physics of dark energy is
intimately connected
to the small-scale ($\sim R_H^{1/3} l_P^{2/3}$) physics of spacetime foam. \\

Alternatively one can 
interpret these long-wavelength quanta as constituents of dark
energy, contributing a more or less uniformly distributed cosmic energy density 
and hence acting as a dynamical effective cosmological constant 
\begin{equation}
\Lambda \sim H^2,  
\label{lambda2}
\end{equation}
a result for the magnitude of $\Lambda$ that will be checked in the next 
section.\\

As a corollary to the above discussion, we can now give a heuristic 
argument \cite{llo04,Arzano,plb} (based on quantum gravity consideration)
on why the Universe cannot contain ordinary matter only.
Start by assuming the Universe (of size $l = R_H$) has only ordinary matter
and hence all information is stored in ordinary matter.  According to 
the statistical mechanics for ordinary matter at temperature $T$, energy scales as
$E \sim l^3 T^4$ and entropy goes as $S \sim l^3 T^3$.  Black hole physics can be
invoked to require $E \lesssim \frac{l}{G} = \frac{l}{l_P^2}$.  
Then it follows that the
entropy $S$ and hence also the number of bits $I$ (or the number of degrees of 
freedom on ordinary matter) are bounded by 
$\lesssim (l / l_P)^{3/2}$.  We can repeat verbatim 
the argument given in section 2 on the relationship
between the bound on the number of degrees of freedom in a region with volume 
$l^3$ and 
$\delta l$, the quantum fluctuation of distance $l$, to conclude that, if 
only ordinary matter exists, $\delta l \gtrsim \left( \frac{l^3}{(l / 
l_P)^{3/2}} \right) ^{1/3} =  l^{1/2} l_P^{1/2}$ which is much greater than
$l^{1/3} l_P^{2/3}$, the result found above from our analysis of the
Salecker-Wigner type of gedanken experiments and implied by the holographic
principle. It is now apparent that ordinary matter contains 
only an amount of information dense enough to map out spacetime
at a level with much coarser spatial resolution. 
Thus, there must be other kinds of matter/energy with which the Universe can map
out its spacetime geometry to a finer spatial accuracy than is possible
with the use of conventional ordinary matter.
We conclude that a dark sector indeed exists in the Universe!
One can draw this conclusion, independent of recent observations of
dark energy and dark matter. We also note that the ($\sim (R_H / l_P)^2$)
bits/``particles" of dark energy
vastly outnumber the ($\sim (R_H / l_P)^{3/2}$) particles of ordinary matter
by an enormously huge factor of $(R_H / l_P)^{1/2} \sim 10^{31}$ for the
present observable universe.\\

\subsection{Dark Energy as Quanta of Infinite Statistics}

According to the holographic spacetime foam model, the constituents of
dark energy are quanta/``particles" with very long wavelengths 
(of the order of Hubble radius $R_H$).
Consider $N \sim (R_H/l_P)^2$ such ``particles'' and let us assume that they 
obey Boltzmann statistics in volume $V \sim R_H^3$ at $T \sim
R_H^{-1}$, the average energy carried by each ``particle".
The partition function $Z_N = (N!)^{-1} (V / \lambda^3)^N$ gives
the entropy of the system $S = N [ln (V / N \lambda^3) + 5/2]$,
with thermal wavelength $\lambda \sim T^{-1} \sim R_H$.
But then $V \sim \lambda^3$, so $S$ becomes negative unless $N \sim 1$
which is equally nonsensical.  A simple solution is to stipulate that
the $N$ inside the log in $S$, i.e, the Gibbs factor
$(N!)^{-1}$ in $Z_N$, must be absent.  (This means that
the N ``particles'' are distinguishable!)
Then the entropy is positive: $S = N[ln (V/ \lambda^3) + 3/2] \sim N$. Now,
the only known consistent statistics in greater than 2 space dimensions
without the Gibbs factor is the quantum Boltzmann statistics, also known as 
infinite statistics. \cite{DHR,greenberg} (See Appendix B for a succinct 
description of this exotic statistics.) Thus we conclude (at least are led to 
speculate) that the ``particles'' constituting dark energy
obey infinite statistics, rather than the familiar Fermi or Bose
statistics. \cite{plb,minic}  
This is the over-riding difference between dark energy and conventional matter. 
Note that here it is the physical non-negativity requirement of entropy for a 
gravitational system that leads to this unexpected conclusion.\\

\section{From Causal-set Theory and Unimodular Gravity to Space-time Foam} 

In this section we will rederive the magnitudes of $\delta l$ (Eq.~(\ref{nvd1}))
and $\Lambda$
(Eq.~(\ref{lambda2})) by using causal-set theory and (generalized) unimodular
gravity. The
causal-set theory \cite{sorkin} stipulates that continuous geometries in
classical gravity should
be replaced by ``causal-sets", the discrete substratum of spacetime. In the
framework of the
causal-set theory, the fluctuation in the number of elements $N$ making up the
set is of the
Poisson type, i.e., $\delta N \sim \sqrt{N}$. For a causal set, the spacetime
volume $V_{st}$
becomes $l_P^4 N$. It follows that
\begin{equation}
\delta V_{st} \sim G \sqrt{V_{st}}.
\label{poisson}
\end{equation}
As in section 2.2, let us consider a spherical volume of radius $l$ over the
amount of time
$T = 2l/c$ it takes light to cross the volume. We want to find the minimum of
$\delta l$;
so $\delta V_{st} \sim T (\delta l)^3 \sim l (\delta l)^3$. With the help of
Eq.~(\ref{poisson})
and $\sqrt{V_{st}} \sim l^2$, we recover 
$\delta l \gtrsim (l l_P^2)^{1/3}$.\\ 

As a check on
Eq.~(\ref{lambda2}), we will make use of the theory of unimodular gravity
\cite{unimod,prl,Bij},
more specifically its generalized action given by 
the Henneaux and Teitelboim action
$ S_{unimod} = - (16 \pi G)^{-1} \int [ \sqrt{g} (R + 2 \Lambda) - 2 \Lambda
\partial_\mu {\mathcal T}^\mu](d^3x)dt. $ 
In this theory, $\Lambda / G$ plays the role of ``momentum" conjugate to the
``coordinate"
$\int d^3x {\mathcal T}_0$ which can be identified as the spacetime volume
$V_{st}$.
Hence the fluctuations of $\Lambda /G$ and $V_{st}$ obey a quantum uncertainty
principle, 
$ \delta V_{st} \! \delta \Lambda/G \sim 1. $  This, together with
Eq.~(\ref{poisson}), yields
$\delta \Lambda \sim V_{st}^{-1/2} \sim R_H^{-2} \sim H^2$, where we have used
$\sim R_H^4$
for the whole spacetime volume $V_{st}$ with $R_H$ being the Hubble radius .
Finally, following Baum \cite{Baum} and Hawking \cite{Hawk},
we can argue\cite{prl} that, in the framework of unimodular gravity, $\Lambda$
vanishes to
the lowest order of approximation (i.e., $\Lambda = 0$ dominates the path
integral of the Euclidean vacuum
functional), and that its first order correction is positive (at least for the
the cosmic
epoch corresponding to redshift $z \stackrel{<}{\sim} 1$.)  We conclude that
$\Lambda \sim + H^2$,
contributing a cosmic energy density $\rho$ given by
$ \rho \! \sim \! \frac{1}{l_P^2 R_H^2}, $ as observed.\\

\section{Dark Matter}

The standard cosmological model, $\Lambda$CDM, has been very successful. But
aside from the
fact that dark matter particles have not been (directly) detected, this model
suffers some noticeable
shortcomings, such as missing satellite problem, core/cusp problem,
too-big-to-fail problem, to name
just a few. \cite{Tulin}
There are also two serious problems that CDM proponents
have to face: CDM theories fail to explain in a natural way \cite{MOND}
the baryonic Tully-Fisher relation (the asymptotic velocity-mass $v^4 \propto M$
relation)
\cite{TF} for galaxies, and
the presence of a universal acceleration scale in galactic (and cluster)
dynamics \cite{MS,Stacy}. 
These apparent shortcomings of $\Lambda$CDM motivated the author and his
collaborators
to construct the Modified Dark Matter (MDM), \cite{Ho1,Ho2}
a phenomenological dark matter model inspired by the consideration of entropy
and gravitation.
Our approach can be traced to the work of Jacobson on gravitational
thermodynamics \cite{Jacob}
and the work of Verlinde on entropic gravity \cite{Verl}.\\

 

\subsection{From Gravitational Thermodynamics /Entropic Gravity to MDM}

Entropy and gravitation come together in Jacobson's idea of gravitational
thermodynamics.
Essentially Jacobson proposes that gravity is simply a consequence of disorder
as quantified
by entropy.  Applying Bekenstein's idea of black hole entropy \cite{BeHa1}  and
Unruh's
formula \cite{Davies,Unruh} for the temperature experienced by an accelerating
body,
Jacobson is able to derive Einstein's equation. His work is instrumental in
inspiring Verlinde's formulation of entropic gravity which is appropriately
generalized
in the construction of Modified Dark Matter.\\

In order to appreciate how important a role entropy and gravitation play, let us
first summarize the
crucial steps in Verlinde's derivation of the canonical Newton's laws. \\

(I) Newton's 2nd law $ \vec{F} = m \vec{a} $:\\ 
(a) Verlinde uses the first law of thermodynamics to
propose the concept of entropic force 
$ F_{entropic} = T \frac{\Delta S}{\Delta x}. $\\
(b) Then he invokes Bekenstein's original arguments concerning the entropy $S$
of black holes:
$ \Delta S = 2\pi k_B \frac{mc}{\hbar} \Delta x $.\\
(c) Finally he applies the formula for the Unruh temperature, $ k_B T =
\frac{\hbar a}{ 2\pi c}, $\, associated
with a uniformly accelerating (Rindler) observer. \\

(II) Newton's law of gravity $a= G M /r^2$:\\
(a) Verlinde considers an imaginary quasi-local (spherical) holographic screen
of area $A=4 \pi r^2$ with temperature $T$.\\
(b) Then he uses equipartition of energy $E= \frac{1}{2} N k_B T$ with $N =
Ac^3/(G \hbar)$ being the total number of
degrees of freedom (bits) on the screen.\\
(c) Finally he applies the Unruh temperature formula and $E= M c^2$. \\

We can now construct MDM by generalizing Verlinde's proposal to de Sitter (dS)
space with positive cosmological
constant $\Lambda$ (like our accelerating universe).  In such a dS space, the
Unruh-Hawking temperature,
as measured by an inertial observer, is $T_{dS} = \frac{1}{2\pi k_B} a_0$ where
$a_0=\sqrt{\Lambda / 3} \sim H$. 
The net temperature as measured by the non-inertial observer \cite{Deser,Jacob2}
(due to some matter sources that cause the acceleration $a$\,)
is \,$\tilde{T}\equiv T_{dS+a}- T_{dS} =\frac{1}{2\pi k_B} [\sqrt{a^2+a_0^2} -
a_0]$.\\

Part (I) of Verlinde's argument can now be generalized to yield the entropic
force (in de Sitter space)
$F_{entropic}= \tilde{T}\, \nabla_x S= m [\sqrt{a^2+a_0^2}-a_0]$.  For $ a \gg
a_0$, we have $F_{entropic}\approx ma$. 
For the small acceleration $a \ll a_0$ regime (where  the galactic rotation
curves are observed to be flat and the
Tully-Fisher relation holds): $ F_{entropic}\approx m \frac{a^2}{2\,a_0}, $  
which, after some algebra, can be shown
to be equal to $F_{Milgrom} \approx m \sqrt{a_N a_c}\,$ the force law proposed
by Milgrom in his theory of
modified Newtonian dynamics (MOND) \cite{MOND} at the galactic scale.  Here we
have identified
$a_0 \approx 2 \pi a_c$,
with the critical galactic acceleration $a_c \sim \sqrt{\Lambda/3} \sim H \sim
10^{-8} cm/s^2$.  Thus
we have correctly predicted the magnitude of $a_c$ (which Milgrom puts in by
hand).  From our perspective,
MOND is a phenomenological consequence of quantum gravity.  But while MOND is
successful in describing
galactic dynamics, it is considerably less so at the cluster and cosmic
scales.\\

Part (II) of Verlinde's argument is straightforwardly generalized to give
$2 \pi k_B \tilde{T} = \frac{G\,\tilde{M}}{r^2}$,
where $\tilde{M} = M + M'$ represents the total mass enclosed within the volume
$V = 4 \pi r^3 / 3$,
with $M'$ being some unknown mass, i.e., dark matter.  It can be checked that
consistency demands 
$M'= \frac{1}{\pi}\,\left(\,\frac{a_0}{a}\,\right)^2\, M$.  It is noteworthy
that dark matter ($M'$) is
related to dark energy (codified in $a_0$) and baryonic matter (M) in MDM.
Succinctly the force law in MDM is given by 
$F_{entropic} = m [\sqrt{a^2+a_0^2}-a_0] = m\,a_N \left[\,1+ (a_0/a)^2/\pi
\right]$.  Recall that, in the
small acceleration $a \ll a_0$ regime, MDM behaves like MOND.  Thus dark matter
($M'$) of the
kind we have in MDM can behave as if there is no (dark) matter but MOND (which
denies the existence
of dark matter); for this reason, initially
\cite{Ho1,Ho2,Ho3} we called our dark matter model ``MONDian Dark Matter" 
with
which Modified Dark Matter shares the acronym MDM. \\

\subsection{Quanta of MDM Obey Infinite Statistics}

It has been known \cite{LB} that the MONDian force
law can be formulated as being governed
by a nonlinear generalization of Poisson's equation which describes the
nonlinear electrostatics embodied in the
Born-Infeld theory.  It is therefore useful to reformulate MDM, via an effective
gravitational dielectric
medium, motivated by the analogy between Coulomb's law in a dielectric medium
and Milgrom's law for MOND. 
Starting from the Born-Infeld theory of electrostatics, we can write the
corresponding gravitational Hamiltonian density
in the form $H_g = \left(\,\sqrt{A^2+A_0^2}-A_0\,\right)/(4 \pi)$ in terms of
the local gravitational fields
$\vec{A}$ and $\vec{A}_0$.
As in the Verlinde approach, let us assume energy equipartition. Then the
effective gravitational Hamiltonian density is equal to
$H_g = \frac{1}{2}\,k_B\, T_{\rm eff}\,$.  The Unruh temperature formula
$T_{\rm eff} = \frac{\hbar}{2\,\pi\,k_B}\, a_{\rm eff}\,$ implies
that the effective acceleration is given by 
$a_{\rm eff}  = \sqrt{A^2+A_0^2}-A_0\,$, which becomes
$a_{\rm eff} = \sqrt{a^2+a_0^2}-a_0\,$  upon the identification (with the
help of the equivalence principle) of the local
accelerations $\vec{a}$ and $\vec{a}_0$ with the local gravitational fields
$\vec{A}$ and $\vec{A}_0$ respectively.
Thus the Born-Infeld inspired force law takes the form of the MDM force law!\\

Next recall that the equipartition theorem in general states that the average of
the Hamiltonian is given by
$\langle H \rangle = - \frac{\partial \log{Z(\beta)}}{\partial \beta}\,$, where
$\beta^{-1} = k_B T$ and $Z$
denotes the partition function. To obtain $\langle H \rangle = \frac{1}{2}
\,k_B\, T$ per degree of freedom,
even for very low temperature, we require $Z$ to be of the Boltzmann form $Z =
\exp(\,-\beta\, H\,)\,$.
But this is precisely what is called the infinite statistics. (See Appendix B.) 
Thus we have shown that
the quanta of MDM (like those of dark energy as shown in section 4) obey
infinite statistics.\cite{Ho3}  \\

\subsection{Observational Tests of MDM}

{\it Observational tests at both the galactic and cluster scales}
\cite{Edmonds1,Edmonds2,Edmonds3}

Since such tests have been described in a long review article \cite{Edmonds3},
here we do not have to
go into details.  Let us just recall that we have found the emergence of a
critical acceleration parameter related to $\Lambda$ in MDM, and it is found in
correlations between
dark matter and baryonic matter in galaxy rotation curves.
The resulting MDM mass profiles are consistent with observational data at
both the galactic and cluster scales.  (We can point out that MDM is more
economical than CDM in fitting
data at the galactic scale, and it is superior to MOND at the cluster
scale.) Logically (and happily as it 
indeed turns out to be the case), the same critical
acceleration appears both in the galactic
and cluster data fits based on MDM.  \\

There is one technical point that is worth mentioning.  It is related to  the
fact that galaxies and clusters
have very different length scales.  Recall that, in our
construction of MDM, we re-interpret acceleration in terms of temperature of the
Unruh-Hawking kind.  Thus,
in principle, the mass profile $M'=
\frac{1}{\pi}\,\left(\,\frac{a_0}{a}\,\right)^2\, M$, fixed by the ratio of the
corresponding Unruh-Hawking temperatures, can be altered due to some physical
effects
associated with a change of scale. For example, in the presence of gravity, the
temperature is not constant in
space at equilibrium.  As a result, it can be modified due to the
Tolman-Ehrenfest effect \cite{Tolman}.  Such an
effect must be incorporated in working out successfully the dark matter density
profiles. \cite{Edmonds2,Edmonds3} \\

{\it MDM and Cosmology}

To apply MDM to cosmology, we must replace $\tilde{M}$ (a non-relativistic
source)  with the
active gravitational (Tolman-Komar) mass (a fully relativistic source).  In that
case, we have 
$\sqrt{a^2+a_0^2}-a_0  = \frac{G\, (\,M(t)+M'(t)\,)}{\tilde{r}^2} + 4 \pi G
\,p\, \tilde{r} -\frac{\Lambda}{3}\,\tilde{r}$,
where $p$ stands for pressure and $\tilde{r}$ is the physical radius.  Then it
can be shown \cite{Ho1}
that the Friedmann equations are recovered.  Note that if we naively use MOND at
the cluster
or cosmic scale, we would be missing the pressure and cosmological constant
terms,
which could be significant. This may explain why MOND doesn't work well at the
cluster
and cosmic scales, whereas MDM works at both the galactic and
cluster scales and is expected to be completely compatible with cosmology.
\cite{Ho1} \\

{\it MDM and Strong Gravitational Lensing}

Let us comment briefly on strong gravitational lensing in the context of MDM and
MOND.
It is known that the critical surface density required for strong lensing is
$\Sigma_c = \, \frac{1}{4 \pi} \, \frac{c H_0}{G} \, F(z_l, z_s)$, with $F
\approx 10$
for typical clusters and background sources at cosmological distances.  Sanders
argued that,
in the deep MOND limit, $\Sigma_{MOND} \approx a_c / G$. \cite{Sanders} 
Recalling that numerically
$a_c \approx c H_0 / 6$, Sanders concluded that MOND cannot
produce strong lensing on its own: $\Sigma_c \approx 5 \Sigma_{MOND}$. 
On the other hand, MDM mass distribution is expected to be sufficient for strong
lensing
since the natural scale for the critical acceleration for MDM is
$a_0 = c H_{0} = 2 \pi a_c \approx 6 a_c$, five to six times that for MOND.
\cite{NgMDM} \\  

{\it MDM as Puffy Dark Matter}

As shown above, MDM quanta obey infinite statistics.  Hence they are extended
(see Appendix B) and
the well-known tools of effective field theory are inadequate.  
How they interact with ordinary matter and how they self-interact remain 
to be investigated. On the other hand, we can heuristically
argue that MDM may enjoy similar properties as DM that are known to have finite
size (and hence, in
a way, extended). One such type of DM is the Puffy DM \cite{Chu}.\\

Collision-less CDM predictions are known to be in tension with small scale
structure observations. 
Self-interacting dark matter (SIDM) models \cite{Tulin} have been proposed to
address these problems
of $\Lambda$CDM; and observations seem to require DM self-scatter with a
cross-section decreasing
with velocity.  Puffy DM naturally satisfies this observational constraint by
having a finite size that is
larger than its Compton wavelength.    It has been shown to be successful
\cite{Chu} in explaining
observations across a wide range of mass scales spanning dwarf galaxies of the
THINGS sample,
low-surface-brightness spiral galaxies and clusters of galaxies including the
Bullet Cluster. It remains to
be seen if MDM enjoys similar successes, but the prospects look promising.\\

 
\section {Turbulence and Spacetime Foam}

In fully developed turbulence in three spatial dimensions, Kolmogorov scaling
specifies the
behavior of $n$-point correlation functions of the fluid velocity.
The scaling \cite{k41} follows from the assumption of constant energy flux,
$\frac{v^2}{t} \sim \varepsilon$, where $v$ stands for the velocity field of the
flow,
and the single length scale $\ell$ is given as
$\ell \sim v\cdot t$.  This implies that $ v \sim (\varepsilon\, \ell)^{1/3}~, $
consistent with the experimentally observed two-point function
$\langle v^i(\ell) v^j(0)\rangle \sim (\varepsilon\, \ell)^{2/3} \delta^{ij}$.\\

In this section we will show that
there are deep similarities between the problem of quantum gravity and
turbulence \cite{jej08}.
The connection between these seemingly disparate fields is provided by the role
of diffeomorphism
symmetry in classical gravity and the volume preserving diffeomorphisms
of classical fluid dynamics.  Furthermore, in the case of irrotational fluids in
three spatial dimensions,
the equation for the fluctuations of the velocity potential can be written in a
geometric
form \cite{unr95} of a harmonic Laplace--Beltrami equation:
$\frac{1}{\sqrt{-g}} \partial_a( \sqrt{-g} g^{ab} \partial_b \varphi) = 0 ~$.
Here, apart from a conformal factor, the effective space time metric has the
canonical ADM form
$ds^2 = \frac{\rho_0}{c} [ c^2 dt^2 - \delta_{ij}(dx^i - v^i dt)(dx^j - v^j
dt)]$, where $c$ is the sound velocity. 
We observe that in this expression for the metric, the velocity of the fluid
$v^i$ plays the role of the shift vector
$N^i$ which is the Lagrange multiplier for the spatial diffeomorphism constraint
(the momentum constraint)
in the canonical Dirac/ADM treatment of Einstein gravity:
$ds^2 = N^2 dt^2 - h_{ij} (dx^i + N^i dt) (dx^j + N^j dt)$. 
Hence in the fluid dynamics context, $N^i \rightarrow v^i$, and a fluctuation of
$v^i$ would imply a fluctuation
of the shift vector. This is possible provided the metric of spacetime
fluctuates, which is a very loose, intuitive,
semi-classical definition of the quantum foam.\\

Next recall length fluctuations $\delta \ell \sim \ell^{1/3} \ell_P^{2/3}$. 
If one defines the velocity as
$v \sim \frac{\delta \ell}{t_c} $,
where the natural characteristic time scale is $t_c \sim \frac{\ell_P}{c} $,
then it follows that
$v \sim c \big(\frac{\ell}{\ell_P}\big)^{1/3} $. It is now obvious that a
Kolmogorov-like scaling \cite{k41} in turbulence
has been obtained, i.e., the velocity scales as $v \sim \ell^{1/3}$ and the
two-point function has the needed
two-thirds power law. 
Since the velocities play the role of the shifts, they describe how the metric
fluctuates at the Planck scale. 
The implication is that at short distances, spacetime is a chaotic and
stochastic fluid in a turbulent
regime \cite{wh55} with the Kolmogorov length $l$. 
This interpretation of the Kolmogorov scaling
in the quantum
gravitational setting implies that the physics of turbulence may help us
understand the quantum
fluctuation phase of strong quantum gravity.\\

\section{Summary and Discussion}

 We have argued that
the laws of physics that determine the precision with which the geometry of
spacetime can be
measured also limit the power of and the amount of information contained in
black hole computers.
Furthermore, the physics of spacetime fluctuations also yields a(n arguably)
successful dark energy
model in terms of an effective positive cosmological constant (related to the
Hubble parameter). 
Then we show that gravitational thermodynamics/ entropic gravity arguments,
generalized to a spacetime (like ours) with dark energy imitating a (positive)
cosmological
constant, lead to a dark matter model which relates dark energy, dark matter,
and ordinary
(baryonic) matter, and is remarkably consistent with observations at both the
galactic and cluster scales.
Lastly we show that turbulence is intimately related to properties of spacetime
foam in the
gravitational context.  These results spanning black holes, computers,
space-time foam, dark energy, dark matter, and turbulence are testimony to the
unity of nature.
They demonstrate the conceptual interconnections of fundamental physics which
makes crucial
(explicit or implicit) use of the concept of entropy and the dynamics of
gravitation. \\

The confluence of entropy and gravitation has produced some rather novel
results.  We would argue
that none is more intriguing than the manifestation
that both dark energy and dark matter 
have their origins in quantum gravity and that
their quanta obey infinite 
statistics while ordinary particles obey either the 
Fermi or Bose statistics. 
This may be the main
difference between the dark sector and ordinary matter.  Furthermore, theories
of ``particles"
obeying infinite statistics are non-local. (See Appendix B).  So it is quite
conceivable that the non-locality encoded in the holographic
principle, a hallmark of quantum gravity, 
is related to this non-locality 
in infinite statistics. (We note that extremal black holes,
another gravitational
system, also obey infinite statistics \cite{stromvolo}.)\\

We conclude this review paper with an observation (perhaps more like a
speculation).  As the gravitational
thermodynamics and entropic gravity ideas have hinted, gravitation may
ultimately be derived from
thermodynamic/entropic arguments.  And if we also take seriously the recent
proposal that spacetime
geometry/gravitation may simply be an emergent phenomenon from quantum
entanglements, as implied by the conjecture ER = EPR \cite{MaSu} , we can
certainly entertain the idea that even
quantum mechanics could be related to thermodynamics in a deep and unfathomable
way. \cite{Adler} If so, then it
follows that thermodynamics, Einstein's ``meta-theory", may hold the key to
formulating as well as
understanding the ultimate physical laws; and reigning supreme will be its
protagonist -- entropy.\\


\vspace{1in}

\noindent
{\bf Acknowledgments}

I am grateful to the US Department of Energy, the Bahnson Fund  and the Kenan
Professors Research Fund of the
University of North Carolina at Chapel Hill for partial financial support during
the years the research
reported in this review article was carried out.  I thank my many collaborators
(see the References)
for stimulating discussions and fruitful collaborations.\\

\newpage

\noindent

{\bf Appendix A: Energy-momentum Fluctuations and Possible Tests of Spacetime
Foam}\\

\noindent
{\it Energy-Momentum Fluctuations}

Just as there are uncertainties in spacetime measurements, there are
also uncertainties in energy-momentum measurements due to
spacetime foam effects. \cite{wigner}.
Imagine sending
a particle of momentum $p$ to probe a certain structure of spatial extent
$l$ so that $p \sim \hbar/l$.
It follows that $\delta p \sim (\hbar/l^2) \delta l$. Spacetime fluctuations
$\delta l \gtrsim l (l_P/l)^{2/3}$ can now be used to give
$\delta p \gtrsim p \left(\frac{p}{m_P c}\right)^{2/3}$, and
$\delta E \gtrsim E \left(\frac{E}{E_P}\right)^{2/3}$,
where $E_P = m_P c^2 \sim 10^{19}$ GeV is the Planck energy.
Consequently
the {\it dispersion relation} is now modified \cite{yjng05} to read
$E^2 - p^2c^2 - \epsilon p^2c^2 \left({pc \over E_P}\right)^{2/3} = m^2c^4$,
for high energies with $E \gg mc^2$, with $\epsilon  \sim 1$.  This modified
dispersion
relation, in turn, leads to a {\it fluctuating speed of light}
\cite{ACetal,yjng05}:
$v = \frac{\partial E}{\partial p} \simeq c \left( 1 +\frac{5}{6} \epsilon
\frac{E^{2/3}}{E_P^{2/3}}\right)$,
which is energy-dependent and fluctuates around c.\\

\noindent
{\it Possible Ways to Test Spacetime Foam}

There have been numerous proposals to detect spacetime foam, involving 
astronomical high-energy gamma ray observations of distant gamma-ray
bursts and distant quasars, gravity-wave interferometers, and
atom interferometers etc.
But when the proper averaging is carried out (even if there is such 
a formalism), now it appears (at least to this author) that the fluctuations are
 perhaps too small to be detectable with the currently 
available experimental and observational techniques.  Nevertheless, let us 
briefly discuss several of the proposals to detect spacetime foam.\\

I. Observing gamma rays from extragalactic sources:

For photons emitted simultaneously from a distant
source, we expect an energy-dependent spread in their arrival times.
So one idea is to look for a noticeable spread in arrival
times for high energy gamma rays from distant gamma ray
bursts (GRB). This proposal was first made by G. Amelino-Camelia
et al. \cite{ACetal} in another context.
But the time-of-flight differences $\delta t$
increase only with the cube root of the
average overall time $t$ of travel ($\delta t \sim t^{1/3} t_P^{2/3}$)
from the gamma ray bursts to our
detector, leading to a time spread too small to be detectable. \cite{yjng05}\\

Another way is to find out if spacetime foam-induced phase incoherence
of light from a distant galaxy or GRB can make the light wave front noticeably 
distorted so as to lose the sharp ring-like interference pattern around 
the galaxy or GRB; \cite{NCvD,lie03}
or to look for halo structures in the interferometric fringes induced by 
fluctuations
in the directions of the wave vector of light from extragalactic sources. \cite{
CNvD}\\

More recently, my collaborators and I \cite{perlman15} showed 
explicitly how wavefront distortions on small scales cause the
image intensity to decay to the point where distant objects become undetectable 
when the
path-length fluctuations become comparable to the wavelength of the radiation.
We noted that detections of quasars at
TeV energies with ground-based Cherenkov
telescopes seem to have ruled out the holographic spacetime foam model
(with $\delta l$ scaling as $l^{1/3} l_P^{2/3}$). But this claim is subject to
some caveats.
For example, my collaborators and I considered only the
instantaneous fluctuations in the distance
between the location of the emission and a given point on the telescope
aperture.
Perhaps one should average over both the huge number of Planck timescales during
the time it takes 
light to propagate through the telescope system, and over the
equally large number of Planck squares across the detector aperture. It is
then possible that the net fluctuations are exceedingly small; but
at the moment, to the best of my knowledge, there is no formalism for carrying
out such averages. \cite{perlman16} \\

II. Measuring the foaminess of spacetime with laser-based interferometers:

For an interferometer with bandwidth centered at frequency $f$, the relevant 
length 
scale characteristic of the noise due to space-time foam is given by 
$l_P^{2/3} (c/f)^{1/3}$. This uncertainty
manifests itself as a displacement noise (in addition to 
noises from other sources) 
that infests the interferometers.
The hope is that modern
gravitational-wave interferometers, through future refinements, may reach
displacement noise level low enough to test a subset of the space-time
foam models. \cite{AC,found}  But this hope is based on
the assumption that spacetime in between the
mirrors in the interferometer fluctuates coherently for all the photons in
the beam.  However the large beam size in LIGO (compared to the Planck scale)
makes such coherence unlikely.\\

\noindent

{\bf Appendix B: Infinite Statistics}\\

For completeness, here we list some of the properties of 
infinite statistics \cite{DHR,greenberg}. 
A Fock realization of infinite statistics 
is given by $a_k a^{\dagger}_l = \delta_{k,l}$.  
This algebra, known as Cuntz algebra, is described by the average of the 
bosonic and fermionic algebras.
Any two states obtained by acting on $|0>$ with creation operators in
different order are orthogonal to each other:\\
$<0|a_{i1}...a_{iN} a^{\dagger}_{jN}...a^{\dagger}_{j1} |0>
= \delta_{i1,j1} ... \delta_{iN,jN}$,
implying that particles obeying infinite statistics are distinguishable.
Accordingly, the partition function is given by
$Z = \Sigma e^{- \beta H}$, without the Gibbs factor.  It is known that,
in infinite statistics, all representations of the particle permutation
group can occur. Theories of particles obeying
infinite statistics are non-local. \cite{fredenhagen,greenberg}
(To be more precise, the fields
associated with infinite statistics are not local, neither in the sense
that their observables commute at spacelike separation nor in the sense
that their observables are pointlike functionals of the fields.)
 In fact, the number operator $n_i$ (which, we recall, satisfies the condition 
$n_i a_j - a_j n_i = - \delta_{i,j} a_j$)
\begin{equation}
n_i = a_i^{\dagger} a_i + \sum_k a_k^{\dagger} a_i^{\dagger} a_i a_k +
\sum_l
\sum_k a_l^{\dagger} a_k^{\dagger} a_i^{\dagger} a_i a_k a_l +
...,  \label{eq:nonloc}
\end{equation}
and Hamiltonian, etc.,
are both nonlocal and nonpolynomial in the field operators.  It is
also known that
TCP theorem and cluster decomposition still hold; and quantum field
theories with infinite statistics remain unitary. \cite{greenberg}

\end{document}